 \documentstyle[11pt,aaspp4]{article}
\def\arcdeg{^\circ}
\cpright{AAS}{1998}
\slugcomment{To appear in The Astrophysical Journal. \copyright 1998
The American Astronomical Society.  All rights reserved.}
\begin{document}
%
%---------------------------------------------------------------------------
%                                TITLE PAGE
%
\title{A Physical Model of Warped Galaxy Disks} 
\author{Kimberly C. B. New\altaffilmark{1}, Joel E. Tohline, Juhan Frank,
and Horst M. V\"ath\altaffilmark{2}}
\affil{Department of Physics \& Astronomy, Louisiana State University, \\
       Baton Rouge, LA  70803-4001}
\altaffiltext{1}{Currently at the Department of Physics \& Atmospheric
Science, Drexel University, Philadelphia, PA 19104}
\altaffiltext{2}{Currently at SAP AG, 69190 Walldorf, Germany}
%
%
%---------------------------------------------------------------------------
%                                ABSTRACT
%
\begin{abstract}

Warped \ion{H}{1} gas layers in the outer regions of spiral galaxies usually display 
a noticeably twisted structure.  This structure almost certainly arises 
primarily as a result of differential precession in the \ion{H}{1} disk as it settles 
toward a preferred orientation in an underlying dark halo potential well 
that is not spherically symmetric.  In an attempt to better understand the 
structure and evolution of these twisted, warped disk structures, we have 
adopted the ``twist--equation'' formalism originally developed by Petterson  
(1977) to study accretion onto compact objects.  Utilizing more 
recent treatments of this formalism, we have generalized the 
twist--equation to allow for the
treatment of non--Keplerian disks and from it have
derived a steady--state structure of twisted disks that develops from free 
precession in a nonspherical, logarithmic halo potential.  
We have used this steady--state solution to produce \ion{H}{1} maps of five
galaxies (M83, NGC 300, NGC 2841, NGC 5033, NGC5055), which match the general
features of the observed maps of these galaxies quite well.  In addition,
the model provides an avenue through which the kinematical viscosity of the
\ion{H}{1} disk and the quadrupole distortion of the dark halo in each galaxy
can be quantified.  This generalized 
equation can also be used to examine the time-evolutionary behavior of 
warped galaxy disks. 

\end{abstract}
     
%---------------------------------------------------------------------------
%                            TEXT OF MANUSCRIPT
%
\section{Introduction}

In simplest terms, spiral galaxy disks can be described as geometrically 
thin, flat, and circular.  We understand that spiral disks are 
geometrically thin because the gas of which they are composed is cold
(the sound speed of the gas is much smaller than its circular orbital velocity);
and they are both circular and flat because, being dissipative, the
gas is fairly efficient at both minimizing out-of-the-plane motions and
radial excursions that would lead to departures from circular orbits.

Describing spiral disks as {\it perfectly} circular and flat is clearly
an oversimplification, however.  In addition to the nonaxisymmetric
structures that are obvious in optical photographs of many spiral disks, 
21-cm maps of the projected velocity fields of spiral disks often reveal
isovelocity contours that are significantly twisted
(\cite{RLW}; \cite{RN78}; \cite{RCC79}; \cite{N80a}, 1980b; \cite{B81};
 \cite{S85}).   Kinematical tilted-ring models have been constructed in 
an effort to explain the presence of such twists in the velocity maps
of \ion{H}{1} disks.  The models indicate that the outer 
regions of many normal spiral disks are significantly warped out of the 
principal plane that is defined by the optically visible, central portion of 
each galaxy.  The line of nodes that defines the intersection of adjacent
rings of gas in these kinematical models also usually must be twisted significantly 
as a function of radius in order to explain the observed contour maps
(for recent reviews, see \cite{Briggs90}; \cite{Bosma91}; 
\cite{CTSC93}, hereafter CTSC).  

It is not particularly surprising that many galaxies are observed to
possess extended, rotationally flattened disks because such structures
appear to be fairly ubiquitous in gravitationally bound, astrophysical
systems.  (There is strong evidence, for example, that rotationally
flattened disks either exist now or have existed in the past
around our sun, individual planets within our solar system,
numerous protostars, the primary star of many mass--exchanging binary
star systems, and active galactic nuclei.)  
What is peculiar about the \ion{H}{1} disks
of many galaxies is that the disks are significantly warped.  
It is not clear why natural dissipative processes similar to those 
which work effectively to minimize
out-of-the-plane motions in stellar or protostellar
``accretion'' disks are unable to suppress warps in the gaseous disks of galaxies.
As Binney (1992) has reviewed, there still is no generally accepted 
dynamical model of spiral galaxies 
that satisfactorily explains either the origin or the current structure of 
warped \ion{H}{1} disks.

Almost thirty years ago, Hunter \& Toomre (1969) examined whether
normal, infinitesimal bending oscillations might be exhibited by
thin, rotating disks of self-gravitating material, permitting them
to sustain coherent warps for more than a Hubble time.
They concluded that ``for any disk whose density tapers sufficiently
gradually to zero near its edge,'' the frequency spectrum of such
oscillations is at least partly continuous and, as a result, coherent
warps cannot be sustained. 
Over the subsequent decade, 
a number of other ideas surfaced to explain the persistence of warps in
galaxies, each one taking advantage of the demonstrated existence of
dark matter halos around galaxies.  As Toomre (1983) has reviewed,
however, models proposing to use the halo as an active agent to
excite warps in otherwise flat disks -- for example, via the Mathieu
instability (Binney 1981) or via a flapping instability (Bertin
\& Mark 1980) -- each present significant difficulties.  
Toomre (1983) and Dekel \& Shlosman (1983) proposed, instead, that
untwisted steady-state warps may be sustained as a result of steady
forcing by a nonspherical, tilted halo.
Building on the early work of Hunter \& Toomre (1969) and 
the idea that forcing by a nonspherical dark halo can influence the dynamics
of the visible disks of galaxies, Sparke (1986) and
Sparke \& Casertano (1988) have shown that
a discrete warping mode can persist
if the disk is sufficiently self-gravitating and if it is embedded
in a nonspherical halo whose equatorial plane is tilted with
respect to the centralmost regions of the disk. 

Adopting this model of galaxy warps, the following evolutionary
picture emerges.  During the galaxy formation process, gas
which falls into a spheroidal dark matter halo generally will
find that its angular momentum vector is tipped at some nonzero
angle, $i$, away from the symmetry axis of the halo.  If the
gas is cold, it will settle into a rotationally flattened disk that 
is tilted at the same angle $i$ with respect to the 
equatorial plane of the halo.  In the centralmost regions of
the galaxy, where the self-gravity of the gas (or, ultimately,
the combined gas/star system) dominates over the gravitational
influence of the halo, the gas will be content to remain in
orbits that preserve this original tilt.  In the outermost
regions of the galaxy where the gravitational field of the
halo dominates, however, the gas should settle into the
halo's equatorial plane.  As Toomre (1983) and Dekel \& Shlosman
(1983) both sketched in their original concept papers, there 
also should be an intermediate region where the gas will
be influenced significantly by the nonspherical gravitational
fields of both the halo and the central gas (or gas/star) disk.

Through their modeling efforts, 
Sparke (1986) and Sparke \& Casertano (1988) confirmed this
earlier suspicion that in the intermediate region, the gas 
can reside in a steady-state, 
``warped disk'' structure that provides a smooth radial
transition between the separate ``flat disk'' orientations of
the inner and outer regions of the galaxy.  Concentrating on
the dynamics of the centralmost and intermediate regions
-- that is, by building models in which there was effectively
no gas in the outermost regions -- Sparke \& Casertano (1988)
showed that, in steady-state, the warped disk exhibits a
straight line of nodes which precesses slowly and coherently
in a direction retrograde to the orbital motion of the gas.
In a time-dependent simulation, furthermore, Hofner \& Sparke
(1994) showed that settling to this steady-state warped disk
structure occurs from the inside, out, and is driven
not by dissipative processes that are similar to those which
are thought to drive settling in most stellar or protostellar
accretion disks but, rather, 
because ``bending waves carry energy associated with transient
disturbances out toward the disk edge.''  
They also showed that, during an evolution as the bending waves
propagate outward through the intermediate region of the disk,  
a twisted structure can develop and persist until the gas
has had sufficient time to settle into the steady-state
(constant line of nodes) configuration. 
Sparke \& Casertano (1988) and Hofner \& Sparke (1994) have demonstrated 
that, with an appropriate choice of parameters, this 
model of disk warping matches well the observed properties of 
several galaxies with warped disks. (See also Kuijken 1991 and
Dubinski \& Kuijken 1995.)

As Hofner \& Sparke (1994) have pointed out, in galaxies 
with extended \ion{H}{1} disks ``the outermost gas cannot be expected
to form part of a coherent warping mode.''  They did not include
normal dissipative forces in their simulations and therefore
were unable to comment on how such forces might influence the
settling process.  In this paper, we examine the disk-settling
process from the other extreme, ignoring the self-gravity of the
gas but introducing an effective kinematical viscosity into the
dynamical equations in order to simulate the effects of dissipative
forces.  Hence, our effort is complementary to the work of Hofner
\& Sparke (1994) and is most relevant to galaxies with
extended \ion{H}{1} disks -- although there is one 
galaxy used for model comparisons (NGC 2841) that is shared by both works. 
We adopt the view that warps in extended \ion{H}{1} disks
which exhibit substantial twists are transient features.
Independent of precisely what physical process
was responsible for initially placing the gas in an orbit
that is inclined to the halo's equatorial plane ({\it e.g.}, gas infall 
at the time of formation or a recent tidal encounter with another
galaxy), the twisted structure 
can be understood as the result of differential precession in the gaseous disk
as it dissipatively settles toward that ``preferred plane.''

In the past, there has been considerable concern (first enunciated by
Kahn \& Woltjer [1959], but reiterated in the
reviews by both Toomre [1983] and Binney [1991]) that 
differential precession will destroy any warped disk structure
on a time scale that is short compared to a Hubble time
and, therefore, that the mechanism we are examining cannot reasonably be
used to explain the persistence of such structures.  
In the outermost regions of \ion{H}{1} disks, however, precession times are
relatively long and, as was first pointed out by
Tubbs \& Sanders (1979), a warped gas layer can persist 
for a Hubble time if the dark halo in which the
disk is embedded deviates only slightly from spherical symmetry.
By modeling carefully the process of disk settling that is driven
by normal dissipative forces and comparing the models to the observed
kinematical properties of galaxies with extended, warped \ion{H}{1} layers, 
we hope to be able to more carefully examine the viability of such models.

Steiman-Cameron \& Durisen (1988, hereafter SCD88) have developed 
this idea rather extensively.
They have adopted a numerical, cloud-fluid
model to simulate the time-dependent evolution of a galaxy
disk that intially is tilted out of the equatorial plane of an underlying,
spheroidal dark halo.  The disk is assumed to be composed of a 
set of annular mass elements, or ``clouds,'' which act like atoms in a viscous
fluid.  The SCD88 model has offered some valuable physical 
insight into the time-dependent settling process 
that is driven by normal dissipative forces and their dynamically generated 
model of a twisted galaxy disk has been surprisingly successful at matching 
the peculiar optical image of one particular galaxy, NGC 4753 (\cite{SCKD92}).  

Our model is analogous to the one developed
by SCD88 but it derives from an analytical prescription of the 
viscous settling process.  More specifically, we employ 
the ``twisted-disk'' equation formalism first
developed by Bardeen \& Petterson (1975) and Petterson (1977, 1978)
to describe the time-dependent settling of a thin, viscous disk in a 
nonspherical dark halo potential.  This is a rather natural formalism to adopt 
because, as numerous kinematical ``tilted-ring'' models have demonstrated, 
warped \ion{H}{1} galaxy disks display a structure that resembles, 
at least qualitatively, the twisted geometry that had once been thought 
to be important in accretion disks which surround certain compact 
stellar objects (Bardeen \& Petterson 1975; see a recent rejuvenation
of this idea put forward by Maloney \& Begelman 1997).  In adapting the model
to galaxy disks, we have replaced the approximate Keplerian gravitational
potential used in earlier accretion disk work with a logarithmic potential
appropriate to galaxy halos.
(Pringle [1992] also recently described how the twisted--disk formalism
may be adapted to galaxies.)
In the limit of
stress-free precession, our model reproduces the analytical prescription of
disk settling first presented by SCD88, but our model is not constrained to 
this limit.  A more general solution to the governing equations
predicts an exponential settling rate that depends on time to the first
power, rather than on time to the third power as has been derived in the 
limit of stress-free precession.  Furthermore, an analytical, steady-state solution
to the governing equations produces a twisted-disk structure that 
is very similar to previously constructed, kinematical models.
We demonstrate that projected surface density maps and radial 
velocity maps derived from our analytical model match
published \ion{H}{1} maps of five well-studied warped disk galaxies 
(M83, NGC 300, NGC 2841, NGC 5033, and NGC 5055) very well. 

\section{The Generalized Twist-Equation}

Bardeen \& Petterson (1975) and Petterson (1978) have written the 
hydrodynamic equations determining the structure
and evolution of nonplanar, thin accretion disks in a ``twisted-coordinate
system.''  In this system, the position $P$ on each ring of radius $r$
is referenced to a local cylindrical coordinate frame $(r, \psi, z^\prime)$
which has been rotated with respect to the cartesian coordinate system of
the central reference ring by the two orientation angles $\gamma$ and $\beta$
(see Fig.\ 1).  For an appropriate choice of the two functions $\gamma\bigl(r
\bigr)$ and $\beta\bigl(r\bigr)$, including specifically the assumption that 
$\beta\bigl(r\bigr) \ll 1$, the equations
separate into a set of the usual hydrodynamic equations for a flat disk
and a pair of coupled ``twist-equations'' governing the orientation of 
the disk.  Because of its ability to describe an apparently complicated,
fully three-dimensional dynamical problem with a relatively simple and elegant
mathematical model, a number of different groups have subsequently also
adopted this formalism.  However, until recently there has been some 
disagreement over what terms may be dropped from the fully three-dimensional,
nonlinear partial differential equations (due to their ``smallness'' relative to other
terms) when deriving the governing twist-equations.
(For a complete discussion of the various simplifications and
assumptions made in deriving this formalism see
Petterson [1978, 1979]; Hatchett, Begelman, \& Sarazin [1981]; and
Papaloizou \& Pringle [1983].)   As the following brief review points out, the 
disagreements have not been over the general {\it form} of the equations but, rather,
over the precise value of certain order-unity coefficients.  Ultimately, as explained
below, we have adopted the derivation presented by Papaloizou \& Pringle (1983).

According to Petterson (1978), the twist-equations take the 
following form:
\begin{eqnarray}
\dot \beta+v_{r}\beta^\prime&=
%&-\frac{2}{3}rv_{r}\biggl[\beta^{\prime\prime}+
&\frac{\nu}{2}C_0 \biggl[\beta^{\prime\prime}+
%\frac{1}{2}\frac{\beta^\prime}{r}-\gamma^{\prime2}\beta\biggr]\;
C_1\frac{\beta^\prime}{r}-\gamma^{\prime2}\beta\biggr]\;
+\;\frac{1}{2\pi
 v_{\psi}}\int_0^{2\pi} F_{T}\cos\psi\,d\psi ,\nonumber  \\ \\ 
\dot \gamma+v_{r}\gamma^\prime&=&
\frac{\nu}{2}C_0 \biggl[\gamma^{\prime\prime}+
C_1\frac{\gamma^\prime}{r}+
\frac{2\gamma^\prime\beta^\prime}{\beta}\biggr]
\;+\;\frac{1}{2\pi\beta
v_{\psi}}\int_0^{2\pi} F_{T}\sin\psi\,d\psi,\nonumber
\label{P78teqs}
\end{eqnarray}
with $C_0 = 2$ and $C_1 = 1/2$. In these expressions, 
dots and primes denote differentiation with respect to time and space, 
respectively, $v_{r}$ and $v_{\psi}$ are the radial and azimuthal
components of the fluid velocity, $\nu$ is the vertically averaged kinematical 
viscosity, and $F_{T}$ is an externally supplied ``twisting force''.

Hatchett, Begelman, \& Sarazin (1981, hereafter HBS) pointed out some 
inconsistencies between Petterson's (1978) 
derivation and the earlier presentation by Bardeen \& Petterson (1975).
Through an independent derivation, they determined that the twist-equations 
do take the form of equation (1), but they concluded that the two coefficients
should have the values $C_0 = 1$ and $C_1 = -1$.  Most significantly, 
HBS showed that the pair of twist-equations can be written as a single, 
{\it complex} twist-equation of the form	
\begin{equation}
\dot w + v_r w^\prime = \frac{\nu}{2}C_0
\biggl(w^{\prime\prime} + C_1\frac{w^\prime}{r}\biggr)
+i{\it L}\;,
\label{HBSteq}
\end{equation}
where the complex variable {\it w} is defined in terms of the angles $\beta$ and 
$\gamma$ as
\begin{equation} 
w\equiv \bigl(\beta\sin\gamma\bigr)-i\bigl(\beta\cos\gamma\bigr)\;,
\end{equation}
and
\begin{equation}
{\it L}\equiv -\frac{1}{2\pi v_{\psi}}\int_0^{2\pi} F_{T}\exp{\Bigl[i\bigl(\gamma+
\psi\bigr)\Bigr]}\,d\psi\;.
\label{Ldef}
\end{equation}
Written in this form, the twist-equation readily submits to analytical solution
for certain driving forces.

Papaloizou and Pringle (1983, hereafter PP) have 
suggested that the HBS derivation also has shortcomings (for example, 
as viscous evolution occurs according to the HBS equations, angular 
momentum does not appear to be conserved globally).
They have derived a set of twist-equations that more consistently 
ties in with traditional $\alpha$-disk models and have demonstrated 
explicitly that their derived equations conserve angular momentum.
As can most easily be deduced from the presentations of Kumar \& 
Pringle (1985) and Pringle (1992), for a disk whose
inner radius goes to zero and whose surface density is independent of
radius, the PP derivation also leads to a complex twist-equation
that takes the form of equation (\ref{HBSteq}).  However, adopting
PP's ``naive approach'' in order to gain physical insight into the 
nature of the problem (i.e., setting Kumar and Pringle's function
$f(\alpha)=\alpha$), one concludes from these three papers that the
proper coefficients are: 
\begin{equation}
C_0 = 1, \eqnum{5a}  
\end{equation}
\begin{equation}
C_1 = \biggl[
2\frac{d\ln (v_{\psi}/r)}{d\ln r} + \frac{d\ln Z}{d\ln r} \biggr],
 \eqnum{5b}
\label{PPcoefs}
\end{equation}
where
\begin{equation}
Z \equiv r^2v_{\psi}.\eqnum{5c}
\end{equation}
\noindent
For a Keplerian disk $v_{\psi} \propto r^{-1/2}$; therefore,
$C_0 = +1$ and $C_1 = -3/2$.
Interestingly, these are the same coefficients that appear in Bardeen \&
Petterson's (1975) earliest presentation of the twist equations.
Henceforth, we will adopt and build upon the twist-equations derived
by PP because it is clear that in a viscous disk
evolution that is governed by the PP twist-equations, 
angular momentum is conserved.

Petterson (1978), HBS, and PP each used an approximately 
Keplerian potential to represent the underlying gravitational potential 
well that governs disk dynamics.  While the Keplerian potential correctly 
describes the gravitational field in which a thin accretion disk surrounding 
a compact object sits, it is inappropriate for a thin galaxy disk sitting in 
the potential well of a nonspherical dark-matter halo.
Utilizing Kumar \& Pringle's (1985) and Pringle's (1992) extensions of
the PP derivation, it is possible to generalize the complex twist-equation 
to incorporate an arbitrary power-law form for the gravitational potential.
Specifically, adopting a rotation law of the form
\setcounter{equation}{5}
\begin{equation}
v_\psi\propto r^{(1-q)},
\label{rotlaw}
\end{equation}
expression (\ref{PPcoefs}) becomes
\begin{equation}
C_1 = \biggl[- 2q + (3-q)\biggr].
\end{equation}  
Furthermore, as Petterson (1978) has pointed out, 
the radial velocity can be expressed in terms of the vertically averaged 
kinematical viscosity as

%%% Complete expression has Sigma gradient and \nu_2. Include here?

\begin{equation}
v_{r}=\nu\frac{r}{v_{\psi}}\frac{\partial}{\partial r}\biggl(\frac{v_{\psi}}
{r}\biggr)\;.
\label{radial velocity}
\end{equation}
Employing our generalized rotation law, we therefore can write
\begin{equation}
v_{r}=-q\frac{\nu}{r}\;.
\label{radvsub}
\end{equation}
Using this expression in conjunction with equations (5a) and (7),
the complex twist-equation (2) takes the form:
\begin{equation}
\dot w=\frac{\nu}{2}  
\biggl[ w^{\prime\prime}+(3-q)\frac{w^\prime}{r}\biggr]
+i{\it L}\;.
\label{BTFteq1}
\end{equation}

%Interestingly, when equation (\ref{radvsub}) is substituted 
%into equation (\ref{BTFteq1})
%the variable $q$ is eliminated.  Hence the single complex twist-equation,
%\begin{equation}
%\dot w=\frac{\nu}{2}\bigl(w^{\prime\prime}+2\frac{w^\prime}{r}\bigr)
%+i{\it L}\;,
%\label{BTFteq2}
%\end{equation}
%serves to describe the time-evolution of twisted disks in any potential well
%that supports a power-law rotation curve.
%(Note:  Had we started with Petterson's [1978] twist-equations, the coefficient of the
%second term in the parentheses would be $\bigl[2q-1\bigr]$ instead of 2
%and such a generalized equation would not have been realized.)

\section{Twisted Disks in a Scale-Free Logarithmic Potential}

As outlined in $\S$ 1, our objective is to examine the evolution of a cold
gaseous disk that is embedded in a spheroidal, dark halo potential.  In
general, we assume that when the gas is initially introduced into the
halo potential well, the orientation of its angular momentum vector
is not aligned with the symmetry axis of the halo.  Because the halo is not 
spherically symmetric, the halo's gravitational field will exert a finite 
``twisting force'' on the gaseous disk; in simplest terms, this field
forces rings
of gas at different radii to precess about the symmetry axis (or symmetry
plane) of the halo.

%The potential of a galaxy disk surrounded by 
In the present context, a spheroidal dark-matter halo
can be satisfactorily represented by a scale-free logarithmic 
potential (cf. \cite{R80}; \cite{SCD88}, 1990).  
This potential is a useful approximation because 
a circular disk orbiting inside such a potential will exhibit a flat rotation
curve (i.e., $q = 1$ and $v_{\psi}$ is independent 
of radius), as seen in real galaxies.
In a halo potential of this type,  the precession frequency
about the symmetry axis of the potential $\omega_{p}$ is
\begin{equation}
\omega_{p}=\pm \frac{3}{4}\eta\frac{v_{\psi}}{r}\;,
\end{equation}
where $\eta$ is a constant which measures the strength of the quadrupole 
distortion of the halo and the minus (plus) sign indicates that the spheriodal
halo is oblate (prolate).

For a twisted disk embedded in a spheroidal potential well, the twisting
force that enters the derived twist equation assumes precisely the same 
algebraic form as the twisting force that Bardeen \& Petterson (1975) 
used when modeling the evolution of a disk orbiting a nonspherical, 
compact stellar object, namely, 
\begin{equation}
F_{T}=2\omega_{p}v_{\psi}\beta\sin\psi\;.
\label{twisting force}
\end{equation}
Hence, from equation (\ref{Ldef}), 
\begin{equation}
{\it L}=\omega_{p}w\;,
\end{equation}
and we deduce that the evolution of a thin galaxy disk in a scale-free logarithmic halo
potential can be described by the expression
\begin{equation}
\dot w=\frac{\nu}{2}
\bigl[w^{\prime\prime}+2\frac{w^\prime}{r}\bigr]
\pm i\biggl(\frac{3}{4}\eta v_{\psi}\biggr)\frac{w}{r}\;.
\label{BTFteq3}
\end{equation}
Unless explicitly stated otherwise, we henceforth will adopt the minus
sign in front of the last term of this expression, thereby assuming that
the halo is {\it oblate} in shape.

Before the character of {\it general} solutions to this complex
ordinary differential equation is examined, it is worthwhile to demonstrate that when
certain limiting physical conditions are imposed, familiar analytical
expressions for the time variation of the inclination angle $\beta(t)$ and
the ``twist'' angle $\gamma(t)$ are derivable from (\ref{BTFteq3}).
This will give us additional confidence that (\ref{BTFteq3}) offers a
valid description of the dynamics of a twisted galaxy disk.  We present,
first, the limit of viscous-free precession, then the limit of stress-free
precession.

\subsection{Viscous-free Precession}

In the limit where viscous effects are completely negligible, $\nu\rightarrow 0$
and equation (\ref{BTFteq3}) reduces to the case of free
precession,
\begin{equation}   
\dot\gamma=-\frac{3}{4}\eta\frac{v_{\psi}}{r}\eqnum{15a}
\end{equation}
with no associated evolutionary damping of orbit inclinations, that is,
\begin{equation}
\dot\beta =0\;.\eqnum{15b}
\end{equation}
\noindent 
Equation (15a) can be integrated to give
\setcounter{equation}{15}
\begin{equation}
\gamma\bigl(r,t\bigr)=\gamma_{0}\bigl(r\bigr)-\biggl(\frac{3}{4}\eta
\frac{v_{\psi}}{r}\biggr)t\;,
\end{equation}
where $\gamma_{0}\bigl(r\bigr)$ prescribes the twisting of the disk at time
$t=0$.

\subsection{Stress-free Precession}

If the viscous stresses due to twisting have negligible effect on
differential precession during the settling process -- a condition
that SCD88 refer to as ``stress-free precession'' -- one may set
$\beta^{\prime\prime}=\gamma^
{\prime\prime}=\beta^\prime=0$ in the governing, complex 
twist-equation.  Under this condition, equation (\ref{BTFteq3})
predicts the following evolutionary behavior:
%\begin{mathletters}
\begin{equation}
\dot \gamma = \nu\frac{\gamma^\prime}{r}-
\frac{3}{4}\eta\frac{v_\psi}{r}\\ \eqnum{17a} 
\end{equation}
\begin{equation}
\dot \beta = -\frac{1}{2}\nu\bigl(\gamma^\prime
\bigr)^{2}\beta. \eqnum{17b}
\end{equation}
%\end{mathletters}
\noindent
This pair of equations matches equations [37a] and [37b] of SCD88
except that the first term on the RHS of equation [17a] is a factor
of 2 larger than the equivalent term in equation [37b] of SCD88.
We suspect that the SCD88 result is in error, but we have been
unable to identify where SCD88 dropped the factor of 2 because their
published derivation was not sufficiently clear.  

If we furthermore adopt the ``stress-free'' precession condition introduced
by SCD88, the first term in equation (17a) is set to zero (hence the 
discrepant factor of 2 is not an issue) and the equation reduces to the 
form of equation (15a) --- hence,
\setcounter{equation}{17}
\begin{equation}
\gamma^\prime=\gamma_{0}^\prime+\biggl(\frac{3}{4}\eta\frac{v_{\psi}}{r^{2}}
\biggr)t\;.
\end{equation}
With $\gamma_{0}^\prime=0$, equation (17b) specifically integrates to give
\begin{equation}
\beta=\beta_0\exp\biggl[-\bigl(t/\tau_{e}\bigr)^3\biggr]\;,
\end{equation}
where
\begin{equation}
\tau_{e}\equiv \biggl[\biggl(\nu/6\biggr)\biggl(\frac{3}{4}\eta\frac
{v_{\psi}}{r^{2}}\biggr)^{2}\biggr]^{-\frac{1}{3}}\;,
\end{equation}
and $\beta_{0}$ is the value of the inclination at $t=0$.  Hence the
analytical settling model presented by SCD88 (see also Steiman-Cameron,
Kormendy, \& Durisen 1992) can be 
straightforwardly derived from our generalized twist-equation.
  
It is not at all clear how widely applicable the SCD88 analytical 
settling model is to the warped \ion{H}{1} disks of normal spiral galaxies
because the simplifying assumptions (e.g., $\beta^{\prime\prime}=
\gamma^{\prime\prime}=\beta^\prime=0$) are quite limiting.  However, with the
analytical twist-equation in hand, we can perform a less restrictive
examination of such systems.

\subsection{General Separable Treatment}

We begin by rewriting equation (\ref{BTFteq3}) in the form
\begin{equation}
\dot w=w^{\prime\prime}+2\frac{w^\prime}{x}-i\frac{w}{x}\;,
\label{BTFteq4}
\end{equation}
where derivatives are now taken with respect to the dimensionless
time and space variables
\begin{eqnarray}
\tau&\equiv & t/t_0\nonumber\\
\;x&\equiv & r/r_0\nonumber,
\end{eqnarray}
and
%\begin{mathletters}
\begin{eqnarray}
t_{0}&\equiv & \frac{8}{9}\nu\bigl(\eta v_{\psi}\bigr)^{-2}\;\label{dimt}\\
\;
r_{0}&\equiv & \frac{2}{3}\nu\bigl(\eta v_{\psi}\bigr)^{-1}\;.
\label{dimr}
\end{eqnarray}
%\end{mathletters}
\noindent 
If we assume that the spatial
and temporal parts of the complex angle $w$ are separable, i.e.,
\begin{equation}
w=T\bigl(\tau\bigr)\cdot\zeta\bigl(x\bigr)\;,
\end{equation}
then
\begin{equation}
T=T_0\exp\bigl[-k^2\tau\bigr],
\end{equation}
where $k$ is in general complex, and $\zeta(x)$ must be a solution of
\begin{equation}
\zeta'' + 2 \frac{\zeta'}{x} + \bigl(k^2 - \frac{i}{x}\bigr)\zeta = 0. 
\end{equation}
The above equation can be transformed by the substitutions $\xi^2 = -4ix$ and
 $\zeta(x) = Z(\sqrt{-4ix})/\sqrt{x}$,
into a form which shows its relationship to Bessel's equation
\def\dd{{\rm d}}
\begin{equation}
\frac{\dd^2 Z(\xi)}{\dd\xi^2} + \frac{1}{\xi}\frac{\dd Z(\xi)}{\dd\xi} +
\biggl(1 - \frac{1}{\xi^2} - \frac{1}{4} k^2\xi^2 \biggr) Z(\xi) = 0 \; .
\label{zeq}
\end{equation}
The character of the solutions and the
spectrum of $k$ are determined by boundary conditions, but some general comments
can be made without reference to specific boundary conditions or initial state.
The fact that one can do so is important here since we do not have a 
well-posed problem in which we know the initial state and the relevant boundary
conditions.  For each $k$ there is a radius $x$ inside which the third term
inside the parantheses in equation (27) may be neglected.  
In this region the solution behaves as $J_1(\xi)$.  
We rule out $Y_1$ because it is not finite at the origin (Abramowicz 
\& Stegun 1972) whereas for plausible initial states $w$ must be finite
everywhere.  In general, the spectrum of $k$ may contain both growing
and decaying modes.  We do not consider growing modes since their
presence would indicate instability.  The fact that one can fit
a number of observational cases with the simpler solution in which only
decaying modes are present may be considered a justification {\it
a posteriori} of this assumption.
In any case, it is clear that if an arbitrary initial state can be written 
as a superposition containing many values of $k$ satisfying appropriate 
boundary conditions, the behaviour in the central region will approach the solution 
with $k=0$. This tendency manifests itself as a function of time first near the 
center and later further out as any contribution with large $Re(k^2)$ dies away 
rapidly. Since the solution with $k=0$ is formally the steady--state solution, 
we would expect it to apply more or less universally to the interior regions of 
twisted disks, no matter what the initial conditions were. The memory of the 
initial conditions is erased inside--out leaving at most an exponentially 
decaying twist in the innermost regions.
 
Because the functions of primary interest to us ultimately are $\beta\bigl(x\bigr)$
and $\gamma\bigl(x\bigr)$, we note that
\begin{eqnarray}
\beta&=&T_0\biggl[{\zeta_R}^2+{\zeta_I}^2\biggr]^{\frac{1}{2}}\exp
\biggl[-Re\bigl(k^2)\tau\biggr], 
\label{beta}\\
\gamma&=&\bigl(\theta + \phi\bigr)+\frac{\pi}{2},
\end{eqnarray}
where:
\begin{eqnarray}
\zeta&=&\zeta_R+i\zeta_I,\\ 
\theta&\equiv & -Im\bigl(k^2\bigr)\tau,\\ 
\phi &\equiv &\tan^{-1}\bigl(\zeta_I/\zeta_R\bigr). 
\label{phi}
\end{eqnarray}

Because $\dot \gamma = \dot \theta = -Im\bigl(k^2\bigr)/t_0$, we recognize also
that for a given $k$, the time--dependent solution of equation (21) presents a
wave (with a ``twisted'' spiral character) that propagates azimuthally through the
disk with a pattern speed
\begin{equation}
{\Omega}_p = -Im\bigl(k^2\bigr)/t_0 = -\frac{8}{9} Im\bigl(k^2\bigr)
\frac{\bigl(\eta v_{\psi}\bigr)^2}{\nu}\;.
\label{omeq}
\end{equation}

\subsection{Steady-State Solution}

Drawing on the HBS discussion, we realize that an analytical solution to the
steady-state problem can be derived.  Specifically, setting 
$\dot w=0$ in equation (\ref{BTFteq4}) or $k^2=0$ in equation (\ref{zeq}), we obtain 
formally
\begin{equation}
w=\frac{1}{\sqrt{x}}Z_{1}\bigl[\sqrt{-4ix}\bigr]\;,
\label{BTFteq5}
\end{equation}
where $Z_{1}$ is any first-order Bessel function of the first kind
(see Gradshteyn \& Ryzhik 1965, \S 8.491, equation 3).  
The physically most
relevant solution appears to be $Z_{1}=J_{1}$ because in this case 
%$\beta\rightarrow 0$ 
$\beta$ is finite as $x\rightarrow 0$ 
and as $x$ increases, $\beta$ increases
monotonically (Abramowicz \& Stegun 1972).
As a point of reference for all subsequent models, then,
we define the steady-state twisted-disk function
\begin{equation}
w_{ss}\equiv \frac{1}{\sqrt{x}}J_{1}\bigl[\sqrt{-4ix}\bigr] = 
\frac{1}{\sqrt{x}}[{\rm ber}_1(2\sqrt{x}) + i {\rm bei}_1(2\sqrt{x})]\; ;
\label{wss}
\end{equation}
where ${\rm ber}_1$ and ${\rm bei}_1$ are the Bessel--real and Bessel-imaginary or 
Kelvin 
functions (see e.g. McLachlan 1941; Abramowicz \& Stegun 1972).
This solution leads to the functions $\beta_{ss}(x)$
and $\gamma_{ss}(x)$ shown in Fig.\ 2. From (\ref{wss}) one can obtain $\beta_{ss}$
and $\gamma_{ss}$ in closed form as follows:
\begin{mathletters}
\begin{eqnarray}
\beta_{ss}  &= \frac{1}{\sqrt{x}} M_1(2\sqrt{x}), \\
\gamma_{ss} &= \theta_1(2\sqrt{x}) + \frac{\pi}{2} ;
\end{eqnarray}
\end{mathletters}
\noindent
where $M_1(x)$ and $\theta_1(x)$ are the modulus and phase functions for
${\rm ber}_1(x)$ and ${\rm bei}_1(x)$.
Over the interval $1 \lesssim x \lesssim 20$, $\beta_{ss}$ and $\gamma_{ss}$ are 
approximated well by the expressions:
\begin{mathletters}
\begin{eqnarray}
\beta_{ss} &\approx & 1 + 1.20\biggl(\frac{x}{10}\biggr)
+ 0.75\biggl(\frac{x}{10}\biggr)^2
+ 1.22\biggl(\frac{x}{10}\biggr)^3,\\
\gamma_{ss} &\approx & \frac{\pi}{4}\biggl[1 + 6.8\biggl(\frac{x}{10}\biggr)
- 3.9\biggl(\frac{x}{10}\biggr)^2
+ 1.63\biggl(\frac{x}{10}\biggr)^3
- 0.28\biggl(\frac{x}{10}\biggr)^4\biggr].
\end{eqnarray}
\end{mathletters}
\noindent
As we shall illustrate presently,
the {\it form} of the function $w_{ss}$ over the interval
$1 \lesssim x \lesssim 20$ defines 
a warped disk model whose features match the observed
properties of a number of individual \ion{H}{1} disks.

\section{Application to Galaxies}
\subsection{Standard Model}

The steady-state solution just derived is formally inappropriate for real galaxies
especially because the outermost regions of real galaxy disks are unlikely to be
described well by time-independent conditions.  However, over the (innermost) regions
of a galaxy disk where differential precession and viscous dissipation have been able
to effect appreciable settling in a Hubble time, we expect the spatial structure of the
disk to assume a form that is very similar to the one portrayed by the function
$w_{ss}\bigl(x\bigr)$.  More specifically, the discussion in \S 3.3 
and equations 
(\ref{beta})-(\ref{phi}) suggest a general time-dependent behavior of the form: 
\begin{mathletters}
\begin{eqnarray}
\beta&\sim&g\vert w_{ss}\vert,
\label{betass}\\
\gamma&\sim&\tan^{-1}\bigl[-Re(w_{ss})/Im(w_{ss})\bigr],
\label{gammass}
\end{eqnarray}
\end{mathletters}
\noindent
where
\begin{equation}
g\sim\beta_0e^{-\sigma\tau}\;,
\eqnum{38c}
\label{f}
\end{equation}
is a spatial constant that describes the amplitude of the warp at a given
time and whose time-evolution follows the above form with $\sigma$ being the
lowest $Re(k^2)$ compatible with boundary conditions.  Again, since we do
not have a well-posed problem, all we can say is that whatever restrictions
the boundary conditions impose on the spectrum of $k$, 
only the lowest $Re(k^2)$ will survive after some time.  When comparing our model
to the properties of real galaxy disks, we will treat $g$ as a free parameter
that measures the overall amplitude of a galaxy's warp 
and not be immediately concerned about the values of $\beta_{0}$, $\sigma$,
or the age of each disk.

The other free parameter that may be adjusted before our twisted disk model
is compared to real galaxy disks is  $x_{max}\equiv r_{max}/r_{0}$, the
radial cutoff to the function $w_{ss}\bigl(x\bigr)$ that will correspond to 
the maximum radial extent $r_{max}$ of a galaxy's \ion{H}{1} disk.
Although $r_{max}$ can be specified for individual galaxies
(see Table 1), the length scale $r_{0}$ as defined in equation (\ref{dimr}) cannot be
specified {\it a priori} because neither of the physical parameters
$\nu$ or $\eta$ is known for individual galaxies.
Hence, initially it was unclear to us what cutoff radius $x_{max}$
({\it i.e.}, what range $0 \leq x \leq x_{max}$) should be adopted
for our model.  Fortunately, from previously published tilted--ring models of
warped \ion{H}{1} disks we were able to determine how rapidly
the warp angle $\beta$ and the twist angle $\gamma$ vary with radius in
real galaxies.  More specifically, we measured the ``pitch--angles"
$d\ln \beta (x)/d\ln x$ and $d\ln \gamma(x)/d\ln x$ in several systems,
then looked for the range(s) of $x$ in our steady--state model over which
both pitch--angle values occurred simultaneously.  We found good matches only
when $x_{max} \sim 10$, that is, galaxy models which match the observations
cannot be produced with values of $x_{max}$ that are orders of magnitude less than
or greater than $10$.
Previously constructed tilted--ring
models also directed the choices of the maximum warping angles
$g$ in our models.

\subsection{Comparison with Specific Galaxies}
By selecting a single value of $x_{max}$ (specifically, $x_{max} = 13$)
and a rather narrow range of warp amplitudes $g$, we have been able to
match many of the qualitative features that are frequently seen in 
the published \ion{H}{1} maps of galaxies.
For example, using the model parameters identified in Table 1, we have
produced the surface brightness maps and radial velocity maps
depicted in Fig.\ 3 for comparison with published maps of the galaxies
M83, NGC300, NGC2841, NGC5033 and NGC5055. 

Figure 3 actually has been pieced together from five separate 
frames of a $>$ 600-frame digital animation sequence in which the viewer 
``flies around'' our model disk, examining it 
from a variety of different lines of sight.  The two angles $i_o$ and $t_o$
uniquely define the orientation of the 
observer's line of sight with respect to the $(x,y,z)$ Cartesian coordinate system (see Fig.\ 
1) that is fixed in the body of the galaxy.  
(When the inclination angle $i_{o} = 0\arcdeg$ or 
$360 \arcdeg,$ the galaxy is being viewed face-on with its angular momentum 
vector pointing 
directly at the observer; when $i_{o} = 90\arcdeg,$ the galaxy is seen edge-on with its 
angular momentum vector pointing down.  When the azimuthal orientation angle $t_{o} = 
0,$ the $+x$ axis is pointing to the right in each image; when $t_{o} = 90\arcdeg,$ 
the $-y$ axis is pointing to the right.)  In each frame of Fig.\ 3, the disk is displayed in 
three 
different ways:  the righthand image is the disk's projected 2D velocity contour map; 
the central image is the disk's projected 2D surface brightness map; and the 
lefthand image presents a 3D rendering of an isodensity surface that encloses virtually the 
entire disk.  Each of the velocity contour maps and surface brightness maps in Fig.\ 3 were 
produced with the 3D radiative transfer routine developed by V\"ath (1994); each 3D 
isodensity
surface was generated using the IDL imaging package.  In the lefthand image of each 
frame, 
the light source reflecting from the disk's 
surface originates from the same position as the viewer's eye.  The five frames displayed in 
Fig.\ 3 were selected from our animation sequence because they identify observer lines of 
sight 
from which our model's 
projected 2D maps match well the published maps of the five indicated galaxies.  

NGC 2841:  To produce the frame of Fig.\ 3 that is identified as NGC 2841, we set 
$g=1.2\times 10^{-2}$ 
in our model disk which corresponds to a maximum warp at the edge of the disk 
of $4.5\arcdeg$.  In Fig.\ 3, the galaxy's angular momentum vector is pointed almost 
directly 
straight up (the disk is tipped slightly so that the viewer is seeing the ``top'' side of the 
galaxy), so velocity contours to the right of the kinematic minor axis are red-shifted and 
contours to the left are blue-shifted.  Our 2D projected surface-brightness and velocity 
maps should be compared, respectively, with the \ion{H}{1} maps published as Figs.\ 7a 
and 7b in Bosma (1981).  Bosma's maps have been reprinted here in Fig.\ 4a.
In order to make the relevant comparisons, the observational maps
have been rotated counterclockwise approximately $125\arcdeg$ in order to properly
orient them with 
respect to our model images.  After this rotation has been made, as our 3D 
rendering
illustrates, the southern edge of our model disk is farther away from the observer than is its
northern edge; this is consistent with the tilt interpretation one derives from an optical
photograph of NGC 2841 and dust obscuration arguments (cf. Plate 14 in Sandage 1961 
and Table 3 of CTSC).  

The velocity contour map of our model of NGC 2841 displays the same gentle, 
counter--clockwise twist as the observed \ion{H}{1} velocity map.  Furthermore,
the north-south extensions seen in Bosma's 
surface-brightness map of NGC 2841 can be identified with the ``leading arm'' 
features that appear in the projected surface-brightness image of our twisted disk 
model.

M83:  To produce the frame of Fig.\ 3 that is identified as M83, we set 
$g=2.2\times 10^{-2}$ in our 
model disk, which corresponds to a maximum warp at the edge of the disk of 
$8.3\arcdeg$.  
The model is oriented so that, in Fig.\ 3, the galaxy's angular momentum vector is pointed 
into the page, almost directly away from the observer. (With this orientation, M83's 
optically
visible spiral arms -- cf. Plate 28 in Sandage 1961 -- are ``trailing'' spiral features.)
The disk is not being viewed 
precisely face-on; as displayed in Fig.\ 3, it is oriented such that, even if the disk were 
perfectly flat, its upper half would be tipped slightly away from the observer and its lower 
half toward the observer.  As a result of this orientation, velocity contours to the right of 
the kinematic minor axis are blue-shifted and contours to the left are red-shifted.  Our 2D 
projected surface-brightness and velocity maps should be compared, respectively, with the 
\ion{H}{1} maps published as Figs.\ 2 and 3 in Rogstad, Lockhart, \& Wright (1974; 
hereafter RLW).  RLW's maps have been reprinted here in Fig.\ 4b.
In order to make the relevant comparisons, the observational maps have been 
rotated counterclockwise approximately $180\arcdeg$ in order to properly orient
them with respect to our model images.

The velocity contour map produced by our twisted disk model of M83
displays most of the broad features that appear in the \ion{H}{1} velocity map published
by RLW.
The surface-brightness map published by RLW displays a pair of  
faint, ``leading spiral'' arms that are not noticeable in the image we have produced in Fig.\ 3 
from our model.  (The features are actually present in our model at a very low amplitude 
level  and can be enhanced somewhat by adjusting the model parameter $g$.  See, for 
example, the tilted--ring model developed by RLW and the model published by CTSC.)  
Analogous features {\it are} present in our surface-brightness image
of NGC 300, presented below. As the 3D isodensity surface of our model of M83 
suggests, these spiral--arm features probably appear in projected surface-brightness
maps of M83 because the galaxy's twisted \ion{H}{1} disk is bending toward 
the observer in the northwestern quadrant and away from the observer in the
southwestern quadrant.

NGC 5033:  To produce the frame of Fig.\ 3 that is identified as NGC 5033, we set 
$g=3.2\times 10^{-2}$ 
in our model disk which corresponds to a maximum warp at the edge of the disk of 
$12\arcdeg$.  In Fig.\ 3, (see also Fig.\ 5 and the related discussion in \S 4.3,
below),
the galaxy's angular momentum vector is pointed down and away 
from the observer, so velocity contours to the right of the kinematic minor axis are 
blue-shifted and contours to the left are red-shifted.  Our 2D projected 
surface-brightness and velocity maps should be compared, respectively, with the 
\ion{H}{1} maps 
published as Figs.\ 4a and 4b in Bosma (1981).  Bosma's maps have been 
reprinted here in Fig.\ 4c.  In order to make the relevant comparisons, 
the observational maps have been rotated counterclockwise approximately 
$100\arcdeg$ in order to properly 
orient them with respect to our model images. 

Both the velocity contour image and the surface-brightness image produced by
our model of NGC 5033 bear a striking resemblance to Bosma's published maps.
The relatively steep density gradients that 
are visible at the northeast and southwest edges of Bosma's surface-brightness maps are, 
according to our model, clearly due to the disk bending toward the observer at one edge 
and away from the observer at the other.  

NGC 5055:  To produce the frame of Fig.\ 3 that is identified as NGC 5055, we set 
$g=5.7\times 10^{-2}$ 
in our model disk which corresponds to a maximum warp at the edge of the disk 
of $21.3\arcdeg$.  This galaxy's angular momentum vector, like that of NGC 5033 (notice 
that $i_{o}$ is the same in these two systems), is pointed down and away from the 
observer, so velocity contours to the right of the kinematic minor axis are blue-shifted and 
contours to the left are red-shifted.  Our 2D projected surface-brightness and velocity maps 
should be compared, respectively, with the \ion{H}{1} maps published as 
Figs.\ 6a and 6b 
in Bosma (1981).  These maps are reprinted in Fig.\ 4d of this manuscript.
In this case, the observational maps did not need to be rotated more than 
$\sim 10\arcdeg$ clockwise in order to make the relevant comparisons.  

Again, the velocity contour image and the surface-brightness image produced by our model
bear a strong resemblance to Bosma's published maps.
Notice, in particular, that from this line-of-sight,
thin arcs are displayed in the upper lefthand and lower righthand regions of our 
model's projected surface-brightness map.  Bosma's \ion{H}{1} 
surface-brightness map definitely displays similar features.  

NGC300:	To produce the frame of Fig.\ 3 that is identified as NGC300, we set 
$g=5.7\times 10^{-2}$ 
in our model disk, which corresponds to a maximum warp at the edge of the 
disk of $21.3\arcdeg$.  The model is oriented so that the galaxy's angular momentum 
vector 
is pointed out of the page, almost directly at the observer. The disk is not being viewed 
precisely face-on; as displayed in Fig.\ 3, it is oriented such that, even if the disk were 
perfectly flat, its upper half would be tipped slightly toward the observer and its lower half 
away from the observer.  As a result of this orientation, velocity contours to the right of the 
kinematic minor axis are blue-shifted and contours to the left are red-shifted.  Our 2D 
projected surface-brightness and velocity maps should be compared, respectively, with the 
\ion{H}{1} maps published as Figs.\ 1 and 2 in Rogstad, Crutcher, \& Chu
(1979; hereafter RCC).  RCC's maps have been reprinted here in Fig.\ 4e.
In order to make the relevant comparisons, the observational maps have been
rotated counterclockwise approximately $160\arcdeg$ in order to properly orient them 
with respect to our model images.
Notice that after this rotation has been made, our 3D rendered image of
NGC300 closely resembles the perspective drawing of RCC's tilted-ring model
(see their Fig.\ 6).  

Notice, first, that the velocity contour image of our model of NGC 300 closely
resembles RCC's published velocity map.  In addition, our
surface-brightness map displays a pair of faint, ``leading 
spiral'' arms that touch the left and right edges of the image shown in Fig.\ 3.  We believe 
that these arms provide an explanation for the faint features that extend to the northwest 
and, particularly, southeast regions of the RCC map.

\subsection{Prograde versus Retrograde Precession}
The model that was used to generate images for Fig.\ 3 was constructed assuming that the
precessional frequency $\omega_{p}$ was negative (see eq. 11).  That is, the underlying
halo was assumed to be {\it oblate} in shape and accordingly, the precessional motions
retrograde with respect to the orbital motion of the gas.  Because the magnitude of the
precessional frequency {\it decreases} with increasing radius in each disk, retrograde
precession results in a physical twist that has a ``leading'' appearance in the 3D,
rendered image of each system.  (In contrast, optically visible spiral arms are generally
thought to be ``trailing'' features.)  Knowing this, one can discern whether the angular
momentum vector of each galaxy is pointing into (out of) the page by simply 
noticing whether
the warped structure exhibits an overall clockwise (counterclockwise) twist.

As CTSC pointed out, an identically good fit to the \ion{H}{1} maps can be obtained with
a disk that exhibits a retrograde twist (i.e., one that settles into a prolate
spheroidal
halo) as with a disk that exhibits a prograde twist.  After switching the sign of $\omega_p$
in eq. (11) to reverse the sense of the twisting in the 3D model, one need only adjust one's
line-of-sight viewing angle according to the prescription   
\begin{eqnarray}
{i_{o}}^\prime&\equiv & \pi-i_o\;\\ \;
{t_{o}}^\prime&\equiv & \pi-t_o\;
\end{eqnarray}
in order to generate projected velocity and surface-density maps
from a ``prolate'' model that are identical to the maps generated from
and ``oblate'' model.  Fig.\ 5b illustrates
what the ``prolate'' model for NGC 5033 looks like once this transformation is employed.

If two quite different 3D disk structures (one with a prograde twist and one
with a retrograde twist) can generate identical 2D maps when projected onto the
sky, how can one decipher which offers the physically more realistic model?
As CTSC pointed out, information gleaned from optical photographs can often 
provide the supplementary
information that is needed to identify which model is more realistic.  For example, one 
must
insist that the angular momentum vector of M83 points away from the observer ({\it into} 
the 
page on Fig.\ 3) if its optical spiral arms are to be interpreted as {\it trailing} features;
hence a warped disk model with a prograde twist (and settling in an {\it oblate} spheriodal
halo) as illustrated in Fig.\ 3 is the preferred solution.  Similar arguments (or ones based
on dust obscuration) lead one to conclude that the disks of NGC 2841, NGC 5055
\footnote{In our judgement, CTSC misidentified NGC 5055 as requiring a 
prolate halo.}, and NGC 300
are also settling in oblate spheriodal halos.  However, for NGC 5033, the retrograde 
twisting
(prolate spheriodal halo) model is clearly preferrable to the prograde twisting model.  This
statement is supported by two arguments:  first, dust obscuration in an 
optical photograph of NGC 5033
(cf. Fig.\ 4a of CTSC and Panel 127 of Sandage \& Bedke 1994) identifies 
the western edge of the galaxy as nearer to the observer than its
eastern edge; second, model 5b must be chosen over 5a if the optical
spiral arms are to be interpreted as trailing features.  Hence, the 3D 
image of NGC 5033 shown in Fig.\ 5b is the correct depiction
of the galaxy as inferred from our model fitting.
 
\subsection{Interpretation}

\subsubsection{Limited Range of Model Parameters}
It should be noted that the animation sequence from which the frames shown in Fig.\ 3 
were 
selected was originally constructed with the expressed intention of flying through lines of 
sight that we previously had determined would produce good fits to M83, NGC 5033 and 
NGC 300; the only intrinsic model parameter that was adjusted during the fly-by sequence 
was the overall warp amplitude $g$.  During our viewing of the fly-by sequence,
we also spotted images that matched the published observational maps of NGC 
2841 
and NGC 
5055.  The frames presented in Fig.\ 3 have been taken directly from the original animation 
sequence; absolutely no attempt has been made to fine-tune the model parameters to fit 
these two additional galaxies separately.  Clearly, then, it is primarily the line-of-sight 
viewing angle to each disk and not the parameters defining the underlying structural 
properties of each disk that had to be adjusted from galaxy to galaxy
in order to match the observations.  Considering its simplicity -- 
in particular, the significantly reduced number of free parameters as compared to 
previous tilted-ring models -- the agreement between the projected maps of our
model and observed \ion{H}{1} maps is excellent.  We conclude, therefore, that the 
function $w_{ss}\bigl(x\bigr)$
and the corresponding expressions for $\beta_{ss}$
and $\gamma_{ss}$ given by Eqs. (36a) and (36b)
provide a good template for matching the observed kinematical structure of 
\ion{H}{1} disks.  
Because a single value of the parameter $x_{max}$ and little variation
in the parameter $g$
are required to achieve a good model fit to five separate galaxies, 
we conclude that these five systems intrinsically have very similar warped disk 
structures.

\subsubsection{Physical Interpretation}
Most importantly, however, since the function $w_{ss}\bigl(x\bigr)$ derives from
a physically reasonable dynamical picture of disk settling, there is an expectation
that our model provides insight into the underlying physical properties of
these systems.  For example, since $x_{max} = 13$ for each galaxy, we conclude
from definition (\ref{dimr}) that in each galaxy, the ratio
\begin{displaymath}
\frac{\nu}{\eta}\approx 0.12v_{\psi}r_{max}\sim 600\,\rm{km\,s}^{-1}
\,\rm{kpc}=2\times10^{29} \rm{cm}^2 \rm{s}^{-1}\;,
\end{displaymath}
where $r_{max}$ is the outer radius of the observed \ion{H}{1} disk.  
(Table 1 shows the values of $\nu/\eta$ calculated individually for
each modeled system.)  Also, if our model is relevant at all, the existence 
of each warped layer indicates that viscous settling has been effective in a
Hubble time ($t\sim {H_0}^{-1}$) and $\tau\sim 1/\sigma$.  
Without a proper understanding of the boundary conditions, we are unable to
determine the correct physical value of $\sigma$.
Assuming for the moment that $\sigma \sim 1$ and setting
$H_0 = 75\,\rm{km\,s}^{-1}\,\rm{Mpc}^{-1}$,
we conclude from definition
(\ref{dimt}) that 
\begin{displaymath}
\frac{\nu}{\eta^2}\sim{v_{\psi}}^2{H_0}^{-1}\sim 600\,\rm{km\,s}^{-1} \rm{Mpc}\;.
\end{displaymath}
These two ratios can only hold if, in each system,
\begin{displaymath}
\nu\sim0.6\,\rm{km\,s}^{-1} \rm{kpc}=2\times10^{26} {\rm{cm}}^2 \rm{s}^{-1}\;
\end{displaymath}
and
\begin{displaymath}
\eta\sim{10}^{-3}\;.
\end{displaymath}
It is worth noting that the values obtained for $\nu$ in all cases
(see Table 1) are
consistent with a rough estimate of the viscosity due to 
dissipational cloud--cloud collisions in the solar neighborhood where 
$\nu \sim 0.25$  km/s kpc (Lynden-Bell \& Pringle 1974). Very modest 
variations in cloud dimensions, r.m.s. velocity and filling factor 
from galaxy to galaxy are sufficient to reproduce the range of values
shown on Table 1.  The fact that reasonable viscosity estimates are obtained
by assuming $\sigma\sim 1$ may be considered as a posteriori justification for
our choice.

\section{Discussion and Conclusions}

Utilizing the formalism originally introduced by Petterson to describe
warped and twisted accretion disk structures in Keplerian potentials, we
have derived a single, complex ODE to describe time dependent settling of
an \ion{H}{1} disk in the logarithmic potential that appears to be typical of normal
spiral galaxies.  Over the interval $1 \lesssim x \lesssim 20$, the analytical function
$w_{ss}\bigl(x\bigr)$ -- derived from an analysis of the steady-state limit
of the general twisted-disk equation -- appears to describe quite accurately the
general warped and twisted structure that is exhibited by a number of galaxy
disks.  It should be noted again that our analysis is based on the assumption
that the warping angle $\beta$ is $\ll 1$.  
For galaxies with larger warps (including
some that we have modeled) this is a simplifying assumption and should be
disregarded in more proper treatments (see Pringle [1992]).

From our model fits we conclude that, quite generally, the effective kinematical
viscosity in these neutral hydrogen disks is $\nu\sim0.6$ km s$^{-1}$ kpc.
That is, the effective Reynolds number in these systems is
\begin{displaymath}
R_e\sim\frac{r_{max} v_{\psi}}{\nu} \sim \frac{x_{max}^2 v_\psi}{r_{max}}
\sim6000\;.
\end{displaymath}
According to equation (\ref{radvsub}), this also implies that the ratio of
the radial inflow velocity of the gas to its orbital velocity is
\begin{displaymath}
\frac{v_r}{v_{\psi}}\approx\frac{1}{R_e}\sim\times{10}^{-4}\;.
\end{displaymath}

This model also provides a mechanism by which the parameter $\eta$ -- the 
quadrupole moment of the underlying dark halo potential well -- can be 
measured in spiral galaxies.  Our fits to five normal spirals with well-studied
warped \ion{H}{1} disks specifically indicate that $\eta\sim{10}^{-3}$.  (This value
can be increased to $\eta\sim{10}^{-2}$, with a corresponding factor of ten 
increase in viscosity, only if the age of these disks is
assumed to be 0.1$\,{H_0}^{-1}$ -- which seems unlikely -- or
$\sigma$ in expression 38c is found to be $\sim 0.1$.)  Hence, we conclude 
that the dark halos in which these warped disks sit are, to quite high 
accuracy, spherically symmetric.  This particular conclusion should not 
come as a surprise because some time ago Tubbs and Sanders (1979) pointed out
that if warped disks are identified as transient structures,
the warp can only be sustained for
a Hubble time if the underlying halo potential is very nearly spherical.
Although we have examined in detail here only the steady-state solution, the
derived time-dependent twisted-disk equation provides a tool that can be 
utilized to model the time-evolution of warped \ion{H}{1} disks without resorting to
elaborate numerical techniques.

The twisted-disk formalism in general -- and the analytical function $w_{ss}\bigl(x\bigr)$
in particular -- provides an avenue through which models of warped \ion{H}{1} disks can 
advance from purely kinematical fits (e.g. tilted-ring models) to dynamical models
based on a reasonable physical model.  In the future, we also expect to use the 
time-dependent form of the twisted-disk equations as a point of comparison for fully 
three-dimensional gas dynamic simulations of \ion{H}{1} disks settling into distorted
halo potentials. 
%---------------------------------------------------------------------------
%                             ACKNOWLEDGEMENTS
%
\acknowledgments

We appreciate the assistance that H. Cohl provided us in rendering the
3D galaxy images shown in Figs.\ 3 and 5, and also would like to thank
Dr. A. Bosma and Dr. D. Rogstad for granting permission to reprint figures from
their earlier papers. We also would like to thank the anonymous referee
for suggestions that resulted in substantial improvements in this 
manuscript.
This work has been supported in part by NASA through grants NAGW-2447
and NAG5-2777, and in part by the U.S. National Science Foundation 
through AST-9528424.

%---------------------------------------------------------------------------
%                             REFERENCES
%

\clearpage
\centerline{\bf{FIGURE CAPTIONS}}

Fig.\  1.-- The twisted coordinate system.  The position $P$ on each
ring of radius $r$ is referenced to a local cylindrical coordinate frame 
$(r, \psi, z^\prime)$ which has been rotated with respect to the cartesian 
coordinate system $(x, y, z)$ of the central reference ring by the two 
orientation angles $\gamma$ and $\beta$. (adapted from Petterson 1977)

Fig.\  2.-- The steady--state behaviors of (a) the inclination angle $\beta$
and (b) the twisting angle $\gamma$ in terms of the dimensionless space
variable $x$.

Fig.\  3.-- Models of the five galaxies NGC2841, M83, NGC5033, NGC5055,
and NGC300, based on the steady--state solution of the complex
twist--equation.  In each frame, the lefthand image is a 3D isodensity
surface which encompasses virtually the entire disk; the central image
is the projected 2D surface brightness map of the disk; and the righthand
image is the projected 2D velocity contour map of the disk.

Fig.\  4.-- Observational \ion{H}{1} maps of five galaxies are 
reprinted here for comparison with the models shown in Fig.\ 3.
As detailed in the text, each of these maps has been rotated in
order to properly orient it with respect to the corresponding
model image.  Observed surface brightness and velocity contour
maps appear on the left and right, respectively, of each  panel
(a)-(e).  (a) Maps of NGC 2841 (Bosma 1981); (b) maps of M83 (RLW);
(c) maps of NGC 5033 (Bosma 1981); (d) maps of NGC 5055 (Bosma 1981);
and (e) maps of NGC 300 (RCC).

Fig.\  5.-- 3D isodensity surfaces for the (a) ``oblate'' 
and (b) ``prolate'' models of NGC5033. 

%
%---------------------------------------------------------------------------
\clearpage

\begin{table}

\caption{Model Parameters}

\bigskip

\begin{tabular} {lcrccrcll}
 $\ $Galaxy  & r$_{\rm max}$\tablenotemark{\rm a} & v$_{\psi}$\tablenotemark{a}$\ \ \ \ 
\ $
 & x$_{\rm max}$ & g & $\beta_{\rm max}$ & $\nu$/$\eta  $ &$\ \ \ \ \ \ $ $\nu$ & $\eta$ 
\\
         & (kpc) & (km s$^{-1}$) & & & $(\arcdeg)\ $ & (km s$^{-1}$ kpc) & (km s$^{-
1}$ kpc) & $10^{-3}$ \\
\tableline
 NGC2841 & 32 & 300$\ \ \ \ $ & 13 & 1.2E-2 &  4.5 & 1100 &$\ \ \ \ $ 1.0  & 0.92 \\
 M83     & 23 & 180$\ \ \ \ $ & 13 & 2.2E-2 &  8.3 &  480 &$\ \ \ \ $ 0.53 & 1.1 \\
 NGC5033 & 24 & 220$\ \ \ \ $ & 13 & 3.2E-2 & 12.0 &  610 &$\ \ \ \ $ 0.58 & 0.94 \\
 NGC5055 & 28 & 213$\ \ \ \ $ & 13 & 5.7E-2 & 21.3 &  690 &$\ \ \ \ $ 0.78 & 1.1 \\
 NGC300  & 15 &  94$\ \ \ \ $ & 13 & 5.7E-2 & 21.3 &  160 &$\ \ \ \ $ 0.22 & 1.4  
\end{tabular}
\tablenotetext{a}{From CTSC}
\end{table}

\clearpage

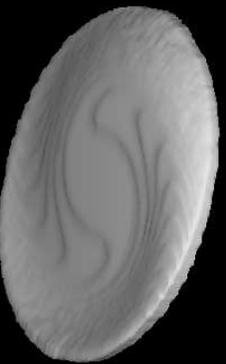
\end{document}